\documentclass[prl,showpacs,twocolumn,floatfix,nofootinbib]{revtex4-1}
\tighten
\usepackage[dvipdfmx]{graphicx}
\usepackage{amsmath,amssymb,natbib,siunitx,color}

\usepackage{color}

\input epsf
\usepackage{graphicx}
\newcommand{\beq}{\begin{equation}}
\newcommand{\beqa}{\begin{eqnarray}}
		  \newcommand{\eeq}{\end{equation}}
\newcommand{\eeqa}{\end{eqnarray}}

\newcommand{\lsim}{\lesssim}
\newcommand{\gsim}{\gtrsim}

\newcommand{\lmk}{\left(}
\newcommand{\rmk}{\right)}

\newcommand{\cp}{{\cal P}}
\newcommand{\cR}{{\cal R}}
\begin{document}

\title{Forecasting  Tidal Disruption Events by  Binary Black Hole Roulettes 
}

\author{Naoki Seto$^1$ and Koutarou Kyutoku$^2$
}
\affiliation{
$^1$Department of Physics, Kyoto University, Kyoto 606-8502, Japan\\
$^2$Interdisciplinary Theoretical Science (iTHES) Research Group, RIKEN,
Wako, Saitama 351-0198, Japan
}

\date{\today}

\begin{abstract}
We discuss the gravitational wave (GW) emission and the orbital evolution of a
 hierarchical triple system composed of an inner binary black hole (BBH)
and an  outer  tertiary.   Depending on the kick velocity at the merger, the merged BBH could tidally disrupt the tertiary.
Even though the fraction of BBH mergers accompanied by such  disruptions is expected to be much 
smaller than unity, the existence of a tertiary and its basic parameters ({\it e.g.} semimajor axis, projected mass) 
can be examined for more than $10^3$ BBHs with
 the space GW detector 
 LISA and its follow-on 
missions. This  allows us to efficiently prescreen the targets for the follow-up searches for the tidal disruption
events (TDEs). 
The TDE probability would be significantly higher for triple systems with  aligned orbital- and spin-angular momenta, 
compared with random configurations.   
\end{abstract}
\pacs{04.30.Tv, 97.60.Lf, 97.80.Kq}

\maketitle

\section{introduction}

The era of gravitational wave astronomy has just started with the surprising discovery of
the GW150914 event by a BBH merger  \cite{ligovirgo2016}.  During its first observational run (O1),
Advanced LIGO detected two BBH mergers \cite{ligovirgo2016-4}. Considering the planned sensitivity improvement and longer
operational time, the total number of BBH detections will increase substantially in the next five years.
  In addition to the ground-based detectors, new windows to the
lower frequency GWs will be opened by  space detectors, and diverse scientific 
possibilities have been actively discussed for  BBHs  at various observational bands \cite{ligovirgo2016-2,sesana2016,kyutoku_seto2016}.

A black hole (BH) is a vacuum solution for the Einstein\rq{}s equation, and is characterized only by its
mass and  spin \cite{wald}. Even though its strong gravity causes many intriguing phenomena, a BH
can be regarded as the simplest astrophysical object, free from complexities of  matter.  
For a merging  BBH, this intrinsic simplicity, in principle, allows us to accurately calculate the kick velocity, 
the  remnant
 mass and the final spin of the merged BBH \cite{pretorius2005,baker_cckm2006,campanelli_lmz2006,lousto2010}.

In this paper, we discuss a triple system composed of an inner BBH and an outer tertiary. 
Because compact binaries are the most promising astrophysical sources of GWs, it
would be meaningful to thoroughly think about the information that we can extract from the  GWs emitted by
 the binaries. 
If the tertiary is a main-sequence  star or a white dwarf, it could be tidally disrupted by the merged
BBH receiving  a suitable kick velocity.   We point out that, only using GW signal from a merging BBH,
 an associated TDE could be forecasted, since the signal could also contain the basic
 information of
the tertiary, such as its semimajor axis and projected mass. 
It is true that the fraction of BBH mergers accompanied by a  TDE is expected to be  much smaller than unity.
However, we will be able to examine the existence of a tertiary individually for totally $\sim 10^3$  nearly monochromatic
BBHs with LISA and for $\sim 10^4$ merging BBHs annually with its follow-on missions. Therefore, we can still expect a chance to
observe a TDE  associated with a BBH merger.

In a recent paper \cite{perets2016} (see also \cite{michaely2016}), mostly for binary systems, 
Perets et al discussed the possibile   TDE (and related processes)  induced by the 
kick at a supernova explosion or even the kick  at a double neutron star merger, referring to \cite{rosswog_dtp2000}.
 Such consecutive events of 
electromagnetic  emissions might be 
observationally advantageous,  since we could  specify the sky position of the subsequent  TDE in advance,
 by identifying the 
preceding explosion.  While it might be interesting  to extend our work by including neutron stars to the inner binaries, 
we only consider the merged BBH as the disrupter, paying special attention to GW observation.   
This is because a BBH is  a very simple system ({\it e.g.}
 for calculating the kick velocity as mentioned earlier). 
 In addition,  electromagnetic emission associated with a BBH merger is likely to be very weak \cite{lyutikov2016}, 
opening the way to a deep follow-up search for the predicted  TDE. 
Inversely, a well localized TDE might also help us to identify or constrain the prompt emissions at the BBH merger.

From GW observation, we can estimate various important parameters for the predicted TDE, including
the mass of the merged BBH, 
its spin magnitude and orientation.
Therefore, once the predicted TDE is actually observed, we might get crucial clues for gaining a further 
understanding of high-energy astrophysical processes, such as  formation and evolution of accretion 
discs and jets.

\section{tidal disruption by merged BBH}

We consider a hierarchical triple system whose inner and outer semimajor axes are $a_b$ and $a_3$.
The inner binary is composed of two BHs with masses $M_1$ and $M_2$, and the outer tertiary is
a star with mass $M_3\ll M_b\equiv M_1+M_2$ and radius $R_3$.   
We apply the subscript \lq\lq{}$b$\rq\rq{} for the inner binary and \lq\lq{}3\rq\rq{} for the 
outer tertiary. 
Below, we set the fiducial mass parameters at 
$M_1=20M_\odot$, $M_2=10M_\odot$ and $M_3=1M_\odot$.

For the alignment of the four angular momentum vectors (two for inner/outer orbits, two for spins of the inner BHs),
 we discuss the following two configurations separately, 1) the optimistic 
configuration: the four vectors are aligned (or anti-aligned), and 2) the random configuration: the 
four vectors are randomly oriented. 
At present, it is still unclear whether the optimistic configuration is a valid  approximation to typical systems.
But, with the large number of BBH observations expected in the forthcoming years,
we will better understand the spin alignment of BBHs \cite{ligovirgo2016,ligovirgo2016-2}. 
For the orbital alignment, some favorable observational
 results   were reported 
recently (see \cite{borkovits2016} for the timing analysis of {\it Kepler} and \cite{ransom2014}
 for a nearly aligned system composed of  a 
millisecond pulsar and two white dwarfs).

For our triple system, the inner and outer orbits are simply assumed to have negligible eccentricities. 
At least for the optimistic configuration,  the inner BBH would  typically 
have a nearly circular orbit, when its GW emission is within the LISA band  or a higher band. Meanwhile, 
the assumption about the outer eccentricity is not essential, as briefly discussed later.

From the Kepler\rq{}s law, the outer orbital velocity $V_3$ and period  $P_3$ are given by 
\beqa
V_3& =&110 (M_b/30M_\odot)^{1/2}(a_3/2{\rm AU})^{-1/2}{\rm km\,s^{-1}},\\
P_3&=&0.52 (M_b/30M_\odot)^{-1/2}(a_3/2{\rm AU})^{3/2}{\rm yr}.
\eeqa

The inner BBH emits GW at the frequency $f=(a_b^3/GM_b)^{1/2}/\pi$ which is twice  the inner orbital
frequency. Due to gravitational radiation reaction, the inner orbit shrinks, while we can reasonably
 assume 
$a_3=$const for a hierarchical triple system.  For our mass parameters, the merger time of the BBH is given as
 \cite{peters_mathews1963}
\beq
T_m=10{\rm Gyr} \lmk  {a_{b}}/{\rm 0.12 AU}\rmk^{4} =177{\rm yr} (f/7{\rm mHz})^{-8/3},
\eeq
and the initial inner separation $a_{b}$ should satisfy the stability condition 
$
a_3/a_{b}\gsim 3.4
$
\cite{eggleton1995}.

Around the merger, the BBH emits strong GWs anisotropically, and, consequently, the merged BH 
receives a kick velocity $V_k$.
The orientation and the magnitude of the kick velocity vector depend strongly on the two masses of the BBH, 
their spins and
orbital phase around the merger. If the two spins are aligned (or counter-aligned) with the orbital angular momentum,
 the kick velocity is on the orbital plane \cite{lousto2010},
 as easily understood from 
the symmetry of the system. With the  mass ratio $M_1/M_2=2$, we have $V_k\sim 160$$\rm km \, s^{-1}$ for 
two spinless
BHs, but $V_k\sim 500$$\rm km \, s^{-1}$ for a certain counter-aligned  spin configuration.  Meanwhile, for 
general spin configuration and mass ratio, the kick velocity can reach $V_k \sim{\rm  5000 km \, s^{-1}}$ 
\cite{zlochower2011}.

For a kick velocity of $V_k\gsim V_3$, the outer orbit is deformed significantly at the BBH merger.
   The tertiary $M_3$ would be tidally disrupted
by the merged BH, if their
closest distance is less than the tidal radius $R_T$. Here we take
$
R_T\equiv R_3 \lmk {2M_{b}}/{M_3} \rmk^{1/3}.
$
At this separation, the characteristic (escape) velocity $V_{esc}=\lmk   {GM_{b}}/{R_T}\rmk^{1/2}$ is explicitly given by 
\beq
V_{esc}=1200\lmk \frac{M_b}{30M_\odot}\rmk^{1/3} \lmk \frac{M_3}{1M_\odot}\rmk^{1/6}
\lmk \frac{R_3}{R_\odot}\rmk^{-1/2}{\rm km\,s^{-1}}.\nonumber
\eeq

For a kick velocity $V_{k}$ satisfying   $V_{3}\ll V_k\ll V_{esc}$, the cross section $\sigma_T$ for the 
close encounter is 
\beq
\sigma_{T}\simeq \pi R_T^2 \lmk \frac{V_{esc}}{V_k} \rmk^2,
\eeq
including gravitational focusing (see {\it e.g.} \cite{perets2016,michaely2016}).  From geometrical consideration, 
this cross section would be a  
conservative estimation  down to $V_k/V_3\sim 1$, and can be applied as  a useful reference to the systems 
analyzed in this paper.
Here we have neglected the effect of the 
radiative mass loss 
(typically $\lsim 5\%$  of $M_b$)  around the merger. 

For the random configuration, the probability of the tidal disruption  is given by 
\beqa
\cp_{ran}=\frac{\sigma_T}{4\pi a_3^2}&\simeq&0.0012 \lmk \frac{R_3}{1R_\odot} \rmk \lmk \frac{a_3}{\rm 2AU} \rmk^{-2} 
\lmk \frac{M_3}{1M_\odot} \rmk^{-1/3} \nonumber\\
& &\times \lmk \frac{M_b}{30M_\odot} \rmk^{4/3} \lmk \frac{V_k}{160 {\rm km\,s^{-1}}} \rmk^{-2}.\label{pr}
\eeqa
 On the other hand, for the optimistic configuration,
the kick is parallel to  the common orbital plane and the probability becomes much higher   
\beqa
\cp_{opt}=\frac{2 \cp_{ran}^{1/2}}{\pi} 
&\simeq&0.022 \lmk \frac{R_3}{1R_\odot} \rmk^{1/2} \lmk \frac{a_3}{\rm 2AU} \rmk^{-1} 
\lmk \frac{M_3}{1M_\odot} \rmk^{-1/6} \nonumber\\
& &\times \lmk \frac{M_b}{30M_\odot} \rmk^{2/3} \lmk \frac{V_k}{160 {\rm km\,s^{-1}}} \rmk^{-1}.\label{po}
\eeqa
The nature is cheating at the BBH roulette via gravitational focusing.
We should also notice that this result is applicable, even if the tilt angle of the outer
orbital plane is less than $O(\cp_{opt})$ radian.
A similar off-plane angle is allowed for the kick velocity that results from slightly tilted spin vectors.

In Eqs. (\ref{pr}) and (\ref{po}), we have $\cp_{ran} \propto \rho_3^{-1/3}$ and $\cp_{opt} \propto \rho_3^{-1/6}$
with the mean density of the tertiary $\rho_3\equiv  (3M_3)/(4\pi R_3^3)$.  For a white dwarf, we typically 
have $\rho_3\sim 10^6{\rm g \,cm^{-3}}$ that is $\sim 10^6$ times higher than a main sequence star of $\sim 1M_\odot$.
Accordingly, the probability of the tidal disruption becomes smaller by a factor of $100$ and $10$, 
respectively for $\cp_{ran}$ and $\cp_{opt}$.

In order to evaluate the final probability that an observed BBH merger is accompanied by  a 
TDE, we also need to know the fraction of BBHs having a tertiary at an appropriate position, in addition to the kinematical 
probabilities (\ref{pr}) and (\ref{po}). 
However, due to our limited understanding of the related astrophysical processes ({\it e.g.} evolution of triple systems),
we currently cannot reliably predict the fraction.  All we are sure is
that the final probability would be much smaller than unity.

\section{ground-based detectors}

Given the expected  small TDE probability associated with a BBH merger,
 it would be crucially advantageous, if we can confirm 
the existence of  a tertiary only using the GW signals from BBHs, and prescreen the suitable targets worth 
for  follow-up TDE searches.

For the fiducial BBH, we only have $T_m\sim 20\,$sec before the merger, when the emitted GW frequency reaches 10Hz. 
This remaining time is much smaller than the reasonable outer orbital period $P_3$. As a result, it would be 
virtually unfeasible to directly detect the tertiary signature, only using ground-based detectors.
Here, one potential approach for more efficiently  searching  a tertiary TDE  is  to concentrate on eccentric BBHs
detected by ground-based detectors. This is based on a circumstantial evidence that the eccentricities of  BBHs
might be caused by tertiaries through 
 the Kozai-Lidov effect (see {\it e.g.} \cite{naoz2016} for a recent review). 
But, such a triple system would typically have random orbital configuration and, resultantly,  
has the smaller TDE probability $\cp_{ran}$.

We should also notice that, with a given detection threshold $\rho_b$, the typical sky localization
 of ground-based detector network for BBHs  would not be better  than
$\sim 10 (\rho_b/10)^{-2} {\rm deg^2}$   \cite{nissanke_kg2013} that is estimated for double neutron stars.   Furthermore the 
expected delay time between a BBH merger and the onset of a potential TDE is 
$\sim 20 (a_3/{\rm 2AU}) (V_3/{\rm 160km\, s^{-1}})^{-1} $ days.  Therefore, even if we can actually detect
electromagnetic wave signals from the  TDE associated with a BBH merger, we might not be confident about the association, 
considering
confusions in a large four-dimensional volume.

When taken the above arguments together, it might be challenging  to carry out a follow-up TDE search in response to the 
BBH mergers observed by ground-based detectors.

\section{signature of a tertiary for space detectors}
In contrast to the ground-based detectors,  space GW detectors can observe BBHs for a longer period of time in the lower 
frequency regime. 
Indeed, as discussed below,  space detectors have potential to find tertiaries and also determine their 
basic parameters, such as the outer semimajor axis $a_3$. In addition, by combining  both 
the ground and space based detectors, 
we can increase the dynamic range of GW observation and more accurately determine the PN parameters that are important
to predict the kick velocity and the radiative mass loss \cite{sesana2016}.

For a hierarchical triple system, the tertiary has the following two celestial mechanical  effects on the inner binary
 \cite{borkovits2016} (see also \cite{meiron2016} for relativistic effects).
One is the dynamical effect that is the corrections to the two-body problem of the inner binary. The other is 
the light travel time effect, corresponding to
the drift of the barycenter of the inner binary.
At $f\gsim 1$mHz relevant for space detectors,  the light travel time effect
 is much larger than the dynamical effect  \cite{borkovits2016}, 
and generates the following phase modulation to the GW from the BBH \cite{seto2008}
\beq
\Psi_3 \sin(2\pi t/P_3+\phi_3).\label{pm}
\eeq
Here the amplitude $\Psi_3$ is given by
\beqa
\Psi_3&=&2\pi f a_3 M_3 \sin I_3 M_b^{-1} c^{-1}\label{amp}\\
&=& 0.66  \lmk \frac{M_3 \sin I_3}{0.5 M_\odot} \rmk \lmk \frac{M_b }{30 M_\odot} \rmk^{-1}
 \lmk \frac{f}{\rm 7 mHz} \rmk \lmk \frac{a_3}{\rm 2AU} \rmk \nonumber
\eeqa
with the inclination angle $I_3$ of the outer orbit relative to the line-of-sight direction. 
The expressions (\ref{pm})-(\ref{amp}) are given for a circular outer orbit. But it is straightforward to
include the outer eccentricity.

Even if the outer orbital frequency $1/P_3$ is not accessible by space detectors,  the phase modulation  (\ref{pm}) is
 up-converted by the carrier GW  frequency $f$.   
For $\Psi_3 \ll 1$, we have two sideband signals at the frequencies $f\pm P_3^{-1}$, and their signal-to-noise ratio 
is given by  
\beq
X_3\sim {\Psi_3 SN_b}/{\sqrt 2} \sim7 \lmk {SN_b}/{30}  \rmk  \lmk {\Psi_3}/{0.33}  \rmk,\label{x3}
\eeq
where $SN_b$ is the signal-to-noise ratio of the intrinsic GWs from the BBH \cite{seto2008} and is assumed to be much larger than unity
 in the present argument. For $\Psi_3\gsim 1$, the quantity $X_3$ can no longer be regarded as the signal-to-noise
ratio of the tertiary, but is still useful for our Fisher matrix analysis below. 

From the phase modulation (\ref{pm}), we can determine the outer orbital elements by
adding the parameters $(\Psi_3,\phi_3,P_3)$ to the matched filtering analysis for the BBH. 
From the Fisher matrix analysis, their  estimation errors are given by
\beqa
\Delta \Psi_3/(\Psi_3)&\sim& \Delta \phi_3\sim 1/X_3,\\
\Delta P_3/P_3&\sim& P_3/T_{obs}X_3,
\eeqa
where $T_{obs}$ is the observational period.
These results are valid for $T_{obs}\gsim 2P_3$ and $|P_3^{-1}-1{\rm yr^{-1}}|\gsim T_{obs}^{-1}$,
since we need to distinguish the tertiary\rq{}s phase modulation from other effects, such as the annual
 motion of the detectors \cite{takahashi_seto2002}.

For the random configuration, we can estimate the outer semimajor axis $a_3$ with error
 $\Delta a_3/a_3\sim {O}(M_3/M_b)$, dominated  by the uncertainty of the 
tertiary mass.  Here, we assumed that the estimation errors for parameters related to the BBH
can be ignored, comparing with those for the tertiary. 

In contrast, for the optimistic configuration, the angle $I_3$ can be determined directly from 
GW polarization measurement, and the mass $M_3$ can be separately estimated with accuracy 
$\Delta M_3/M_3\sim
1/X_3$.  We also have $\Delta a_3/a_3\sim { O}[(M_3/M_b)/X_3]$. But for a nearly face-on geometry 
(potentially relevant for relativistic emissions  towards the spin direction of the merged BBH),
we have $\sin I_3\sim0$ and the tertiary might not be identified. 

Both for the random and optimistic configurations, around the BBH merger,  
we can predict the outer orbital phase 
$\phi_3(t)\equiv 2\pi t/P_3+\phi_3$ with  the accuracy of   ${ O}(1/X_3)$, 
so that a TDE might be forecasted with high confidence as discussed below.

\section{forecasting TDEs}

Using the expressions derived so far, we now discuss the prospects for forecasting TDEs with space 
interferometers (hereafter assuming the fiducial masses for the BBHs). 

While the sensitivity of LISA has not been fixed yet, its relatively 
sensitive version \cite{klein_etal2016} can detect $\sim 10^3 (\cR/100 {\rm Gpc^{-3}yr^{-1}})$ BBHs,
around its optimal frequency $\sim 7$mHz, up to distance $\sim 200$Mpc in the observational period 
$T_{obs}\sim5$yr \cite{kyutoku_seto2016}. 
Here $\cR$ is the comoving merger rate of BBHs
 and the observationally inferred value after LIGO O1 is 
9-240 $\rm Gpc^{-3} yr^{-1}$ \cite{ligovirgo2016-4}.

For a BBH at 7mHz with $SN_b\sim15$, LISA can detect a tertiary, for example,  at $a_3\sim 4$AU with 
$M_3 \sim 0.3M_\odot$ in $T_{obs}\sim 5$yr.  Therefore, LISA would be a powerful tool to make a census of triple systems
including inner BBHs.
However, around the optimal frequency $\sim7$mHz, the BBHs
 would be 
nearly monochromatic with $T_m\gg T_{obs}$ (see Eq. (3)).
Therefore, for a TDE forecast within a desirable time ({\it e.g.} $\lsim 10\,$yr), 
 we need to find merging BBHs at higher frequencies.
But, 
these merging ones would constitute a minor fraction among the  BBHs detected with LISA \cite{kyutoku_seto2016}.

For forecasting TDEs, the preferable frequency regime is just  between the LIGO band and the LISA band, 
and is planned to
be explored by the follow-on missions to LISA, such as BBO \cite{cutler_holz2009} or DECIGO \cite{seto_kn2001}.
For example, with the proposed sensitivity of BBO \cite{cutler_holz2009}, a BBH  
at $z=0.5$ can be detected with the averaged signal-to-noise ratio of
 990. 
This total signal-to-noise ratio  mainly comes from its optimal band $\sim 0.3$Hz.  But, for 
identifying a tertiary of 
$P_3={O}(1{\rm yr})$, we need to analyze the GW signals from the lower frequency part satisfying the condition 
for the remaining time
$T_m(f)\gsim 2 P_3$.
In this regard, if we limit the signal integration only between  $T_m=5$yr  and 1yr 
(corresponding to the observed frequencies $f=0.021$Hz and $0.038$Hz),
 the  signal-to-noise ratio decreases down to 53.2 from the total value 990. 
 From  this partial  signal we can detect a 1$M_\odot$ tertiary at $a_3=4$AU  with $X_3\sim 300$.
Then, for the optimistic configuration, we can figure out the outer orbital phase $\phi_3(t)$ around the BBH merger 
with the accuracy of ${ O}(10^{-2})$ radian.
If the kick velocity vector can be predicted with  sufficient accuracy using the inspiral/merger waves,
 we can  examine the two-body problem for the merged BBH and the tertiary.  This allows us to outwit the BBH roulette
beyond the simple evaluation (\ref{po}), and, furthermore, predict the onset time of the TDE. 
Even for the random configuration, using the observed time
of the TDE, we can estimate the kick velocity and also determine the orientation of the outer orbit with the predicted kick velocity, at least in principle.
We leave these studies as a future work. But, we should  notice that the prediction for the 
kick velocity vector is much easier for the optimistic configuration.  This is because we only need to deal with
 the spin component perpendicular to the orbital plane, and also  
 the inspiral waveform is sensitive mostly to this  component, compared with 
the on-plane components (see {\it e.g.} \cite{poisson1995,ajith2011}).

The expected merger rate of BBHs within $z=0.5$ is $\sim  10^4 (\cR/100 {\rm Gpc^{-3}yr^{-1}})$ per year.
Here, importantly for electromagnetic-wave observation, a BBH at $z=0.5$ can be localized 
by BBO within an error box of $\sim 0.1 \rm arcsec^2$. Additionally using the estimated  distance to 
the BBH, it is likely that we can uniquely specify the host galaxy of the TDE candidate in advance
\cite{cutler_holz2009}.

Observational signatures of TDEs caused by a stellar-mass BH is not understood (see \cite{perets2016} for discussions). Many X-ray transients have recently been identified as TDEs caused by supermassive BHs as well as optical and UV transients \cite{auchett2016}. Because the tidal radius is smaller by more than an order of magnitude for stellar-mass BHs, the emission from the disk or outflow will be primarily bright in X-rays for TDEs associated with our BBH roulette. The minimum fallback time will be shorter for stellar-mass BHs than for supermassive BHs compared at the tidal radius \cite{perets2016}. Thus, the peak luminosity should be higher for the former. Taking also the high localization accuracy by space GW detectors and the low rate of TDEs by supermassive BHs, the confusion may not be severe. In addition, the minimum fallback time and accretion time are both likely to be shorter than the expected travel time, $O(10)$ days, of the merged BBH to the tertiary, and therefore the electromagnetic emission is safely assumed to be coincident with the encounter. Late-time light curves will follow the power law with a canonical index of $\approx -5/3$. This feature will be useful to distinguish TDEs from other transient phenomena. However, the expected large inclination angle, $\sim I_3$ (for the optimistic configuration), will degrade the prospect for observing nonthermal emission from possible relativistic jets. Still, if the jet is not extremely relativistic, it can be observed and may allow us to explore the environment surrounding the triple system, particularly in radio bands.  Observations of forecasted TDEs in the BBH roulette will deepen our understanding of disruption physics by widening the parameter space of TDEs (see also, e.g., \cite{krolik2011,ioka2016}  for disruption of white dwarfs by intermediate-mass BHs and \cite{stephens2011,rosswog2013} for neutron stars by stellar-mass BHs).

The primary goal of BBO/DECIGO is the direct detection of the GW background from inflation \cite{cutler_holz2009,seto_kn2001}. But, in terms of the 
energy density $\Omega_{GW}$, 
 the
primordial signal is expected to be more than $10^5$ times smaller than the foreground GWs produced by cosmological
compact binaries, such as BBHs or double neutron stars \cite{cutler_holz2009}. Therefore, it is essential to identify   the individual 
chirping binaries around 0.1Hz, and remove their contributions from the data streams of detectors.  
Here, various potential 
astrophysical effects should be carefully examined to reduce the residual noise as small as possible. 
Therefore, the proposed tertiary search for a large number of merging BBHs will be conducted as a byproduct 
for realizing the primary goal of BBO/DECIGO.

\begin{acknowledgments}
We would like to thank Kunihito Ioka and Hiroyuki Nakano for helpful discussions.
 This work is supported by JSPS Kakenhi Grant-in-Aid for Research
 Activity Start-up (No.~15H06857), for Scientific Research
 (No.~15K65075) and for Scientific Research on Innovative Areas
 (No.~24103006). Koutarou Kyutoku is supported by the RIKEN iTHES
 project.
\end{acknowledgments}


\begin{thebibliography}{34}%
\makeatletter
\providecommand \@ifxundefined [1]{%
 \@ifx{#1\undefined}
}%
\providecommand \@ifnum [1]{%
 \ifnum #1\expandafter \@firstoftwo
 \else \expandafter \@secondoftwo
 \fi
}%
\providecommand \@ifx [1]{%
 \ifx #1\expandafter \@firstoftwo
 \else \expandafter \@secondoftwo
 \fi
}%
\providecommand \natexlab [1]{#1}%
\providecommand \enquote  [1]{``#1''}%
\providecommand \bibnamefont  [1]{#1}%
\providecommand \bibfnamefont [1]{#1}%
\providecommand \citenamefont [1]{#1}%
\providecommand \href@noop [0]{\@secondoftwo}%
\providecommand \href [0]{\begingroup \@sanitize@url \@href}%
\providecommand \@href[1]{\@@startlink{#1}\@@href}%
\providecommand \@@href[1]{\endgroup#1\@@endlink}%
\providecommand \@sanitize@url [0]{\catcode `\\12\catcode `\$12\catcode
  `\&12\catcode `\#12\catcode `\^12\catcode `\_12\catcode `\%12\relax}%
\providecommand \@@startlink[1]{}%
\providecommand \@@endlink[0]{}%
\providecommand \url  [0]{\begingroup\@sanitize@url \@url }%
\providecommand \@url [1]{\endgroup\@href {#1}{\urlprefix }}%
\providecommand \urlprefix  [0]{URL }%
\providecommand \Eprint [0]{\href }%
\providecommand \doibase [0]{http://dx.doi.org/}%
\providecommand \selectlanguage [0]{\@gobble}%
\providecommand \bibinfo  [0]{\@secondoftwo}%
\providecommand \bibfield  [0]{\@secondoftwo}%
\providecommand \translation [1]{[#1]}%
\providecommand \BibitemOpen [0]{}%
\providecommand \bibitemStop [0]{}%
\providecommand \bibitemNoStop [0]{.\EOS\space}%
\providecommand \EOS [0]{\spacefactor3000\relax}%
\providecommand \BibitemShut  [1]{\csname bibitem#1\endcsname}%
\let\auto@bib@innerbib\@empty
\bibitem [{\citenamefont {{Abbott}}\ \emph
  {et~al.}(2016{\natexlab{a}})\citenamefont {{Abbott}}, \citenamefont
  {{Abbott}}, \citenamefont {{Abbott}}, \citenamefont {{Abernathy}},
  \citenamefont {{Acernese}}, \citenamefont {{Ackley}}, \citenamefont
  {{Adams}}, \citenamefont {{Adams}}, \citenamefont {{Addesso}}, \citenamefont
  {{Adhikari}},\ and\ \citenamefont {et~al.}}]{ligovirgo2016}%
  \BibitemOpen
  \bibfield  {author} {\bibinfo {author} {\bibfnamefont {B.~P.}\ \bibnamefont
  {{Abbott}}}, \bibinfo {author} {\bibfnamefont {R.}~\bibnamefont {{Abbott}}},
  \bibinfo {author} {\bibfnamefont {T.~D.}\ \bibnamefont {{Abbott}}}, \bibinfo
  {author} {\bibfnamefont {M.~R.}\ \bibnamefont {{Abernathy}}}, \bibinfo
  {author} {\bibfnamefont {F.}~\bibnamefont {{Acernese}}}, \bibinfo {author}
  {\bibfnamefont {K.}~\bibnamefont {{Ackley}}}, \bibinfo {author}
  {\bibfnamefont {C.}~\bibnamefont {{Adams}}}, \bibinfo {author} {\bibfnamefont
  {T.}~\bibnamefont {{Adams}}}, \bibinfo {author} {\bibfnamefont
  {P.}~\bibnamefont {{Addesso}}}, \bibinfo {author} {\bibfnamefont {R.~X.}\
  \bibnamefont {{Adhikari}}}, \ and\ \bibinfo {author} {\bibnamefont
  {et~al.}},\ }\href {\doibase 10.1103/PhysRevLett.116.061102} {\bibfield
  {journal} {\bibinfo  {journal} {Phys. Rev. Lett.}\ }\textbf {\bibinfo
  {volume} {116}},\ \bibinfo {pages} {061102} (\bibinfo {year}
  {2016}{\natexlab{a}})}\BibitemShut {NoStop}%
\bibitem [{\citenamefont {{Abbott}}\ \emph
  {et~al.}(2016{\natexlab{b}})\citenamefont {{Abbott}}, \citenamefont
  {{Abbott}}, \citenamefont {{Abbott}}, \citenamefont {{Abernathy}},
  \citenamefont {{Acernese}}, \citenamefont {{Ackley}}, \citenamefont
  {{Adams}}, \citenamefont {{Adams}}, \citenamefont {{Addesso}}, \citenamefont
  {{Adhikari}},\ and\ \citenamefont {et~al.}}]{ligovirgo2016-4}%
  \BibitemOpen
  \bibfield  {author} {\bibinfo {author} {\bibfnamefont {B.~P.}\ \bibnamefont
  {{Abbott}}}, \bibinfo {author} {\bibfnamefont {R.}~\bibnamefont {{Abbott}}},
  \bibinfo {author} {\bibfnamefont {T.~D.}\ \bibnamefont {{Abbott}}}, \bibinfo
  {author} {\bibfnamefont {M.~R.}\ \bibnamefont {{Abernathy}}}, \bibinfo
  {author} {\bibfnamefont {F.}~\bibnamefont {{Acernese}}}, \bibinfo {author}
  {\bibfnamefont {K.}~\bibnamefont {{Ackley}}}, \bibinfo {author}
  {\bibfnamefont {C.}~\bibnamefont {{Adams}}}, \bibinfo {author} {\bibfnamefont
  {T.}~\bibnamefont {{Adams}}}, \bibinfo {author} {\bibfnamefont
  {P.}~\bibnamefont {{Addesso}}}, \bibinfo {author} {\bibfnamefont {R.~X.}\
  \bibnamefont {{Adhikari}}}, \ and\ \bibinfo {author} {\bibnamefont
  {et~al.}},\ }\href {\doibase 10.1103/PhysRevLett.116.241103} {\bibfield
  {journal} {\bibinfo  {journal} {Phys. Rev. Lett}\ }\textbf {\bibinfo {volume}
  {116}},\ \bibinfo {pages} {241103} (\bibinfo {year}
  {2016}{\natexlab{b}})}\BibitemShut {NoStop}%
\bibitem [{\citenamefont {{Abbott}}\ \emph
  {et~al.}(2016{\natexlab{c}})\citenamefont {{Abbott}}, \citenamefont
  {{Abbott}}, \citenamefont {{Abbott}}, \citenamefont {{Abernathy}},
  \citenamefont {{Acernese}}, \citenamefont {{Ackley}}, \citenamefont
  {{Adams}}, \citenamefont {{Adams}}, \citenamefont {{Addesso}}, \citenamefont
  {{Adhikari}},\ and\ \citenamefont {et~al.}}]{ligovirgo2016-2}%
  \BibitemOpen
  \bibfield  {author} {\bibinfo {author} {\bibfnamefont {B.~P.}\ \bibnamefont
  {{Abbott}}}, \bibinfo {author} {\bibfnamefont {R.}~\bibnamefont {{Abbott}}},
  \bibinfo {author} {\bibfnamefont {T.~D.}\ \bibnamefont {{Abbott}}}, \bibinfo
  {author} {\bibfnamefont {M.~R.}\ \bibnamefont {{Abernathy}}}, \bibinfo
  {author} {\bibfnamefont {F.}~\bibnamefont {{Acernese}}}, \bibinfo {author}
  {\bibfnamefont {K.}~\bibnamefont {{Ackley}}}, \bibinfo {author}
  {\bibfnamefont {C.}~\bibnamefont {{Adams}}}, \bibinfo {author} {\bibfnamefont
  {T.}~\bibnamefont {{Adams}}}, \bibinfo {author} {\bibfnamefont
  {P.}~\bibnamefont {{Addesso}}}, \bibinfo {author} {\bibfnamefont {R.~X.}\
  \bibnamefont {{Adhikari}}}, \ and\ \bibinfo {author} {\bibnamefont
  {et~al.}},\ }\href {\doibase 10.3847/2041-8205/818/2/L22} {\bibfield
  {journal} {\bibinfo  {journal} {Astrophys. J.}\ }\textbf {\bibinfo {volume}
  {818}},\ \bibinfo {pages} {L22} (\bibinfo {year}
  {2016}{\natexlab{c}})}\BibitemShut {NoStop}%
\bibitem [{\citenamefont {Sesana}(2016)}]{sesana2016}%
  \BibitemOpen
  \bibfield  {author} {\bibinfo {author} {\bibfnamefont {A.}~\bibnamefont
  {Sesana}},\ }\href {\doibase 10.1103/PhysRevLett.116.231102} {\bibfield
  {journal} {\bibinfo  {journal} {Phys. Rev. Lett.}\ }\textbf {\bibinfo
  {volume} {116}},\ \bibinfo {pages} {231102} (\bibinfo {year}
  {2016})}\BibitemShut {NoStop}%
\bibitem [{\citenamefont {Kyutoku}\ and\ \citenamefont
  {Seto}(2016)}]{kyutoku_seto2016}%
  \BibitemOpen
  \bibfield  {author} {\bibinfo {author} {\bibfnamefont {K.}~\bibnamefont
  {Kyutoku}}\ and\ \bibinfo {author} {\bibfnamefont {N.}~\bibnamefont {Seto}},\
  }\href {\doibase 10.1093/mnras/stw1767} {\bibfield  {journal} {\bibinfo
  {journal} {Mon. Not. R. Astron. Soc.}\ }\textbf {\bibinfo {volume} {462}},\
  \bibinfo {pages} {2177} (\bibinfo {year} {2016})}\BibitemShut {NoStop}%
\bibitem [{\citenamefont {Wald}(1984)}]{wald}%
  \BibitemOpen
  \bibfield  {author} {\bibinfo {author} {\bibfnamefont {R.~M.}\ \bibnamefont
  {Wald}},\ }\href@noop {} {\emph {\bibinfo {title} {General relativity}}}\
  (\bibinfo  {publisher} {The university of Chicago press},\ \bibinfo {year}
  {1984})\BibitemShut {NoStop}%
\bibitem [{\citenamefont {Pretorius}(2005)}]{pretorius2005}%
  \BibitemOpen
  \bibfield  {author} {\bibinfo {author} {\bibfnamefont {F.}~\bibnamefont
  {Pretorius}},\ }\href {\doibase 10.1103/PhysRevLett.95.121101} {\bibfield
  {journal} {\bibinfo  {journal} {Phys. Rev. Lett.}\ }\textbf {\bibinfo
  {volume} {95}},\ \bibinfo {pages} {121101} (\bibinfo {year}
  {2005})}\BibitemShut {NoStop}%
\bibitem [{\citenamefont {Baker}\ \emph {et~al.}(2006)\citenamefont {Baker},
  \citenamefont {Centrella}, \citenamefont {Choi}, \citenamefont {Koppitz},\
  and\ \citenamefont {van Meter}}]{baker_cckm2006}%
  \BibitemOpen
  \bibfield  {author} {\bibinfo {author} {\bibfnamefont {J.~G.}\ \bibnamefont
  {Baker}}, \bibinfo {author} {\bibfnamefont {J.}~\bibnamefont {Centrella}},
  \bibinfo {author} {\bibfnamefont {D.-I.}\ \bibnamefont {Choi}}, \bibinfo
  {author} {\bibfnamefont {M.}~\bibnamefont {Koppitz}}, \ and\ \bibinfo
  {author} {\bibfnamefont {J.}~\bibnamefont {van Meter}},\ }\href {\doibase
  10.1103/PhysRevLett.96.111102} {\bibfield  {journal} {\bibinfo  {journal}
  {Phys. Rev. Lett.}\ }\textbf {\bibinfo {volume} {96}},\ \bibinfo {pages}
  {111102} (\bibinfo {year} {2006})}\BibitemShut {NoStop}%
\bibitem [{\citenamefont {Campanelli}\ \emph {et~al.}(2006)\citenamefont
  {Campanelli}, \citenamefont {Lousto}, \citenamefont {Marronetti},\ and\
  \citenamefont {Zlochower}}]{campanelli_lmz2006}%
  \BibitemOpen
  \bibfield  {author} {\bibinfo {author} {\bibfnamefont {M.}~\bibnamefont
  {Campanelli}}, \bibinfo {author} {\bibfnamefont {C.~O.}\ \bibnamefont
  {Lousto}}, \bibinfo {author} {\bibfnamefont {P.}~\bibnamefont {Marronetti}},
  \ and\ \bibinfo {author} {\bibfnamefont {Y.}~\bibnamefont {Zlochower}},\
  }\href {\doibase 10.1103/PhysRevLett.96.111101} {\bibfield  {journal}
  {\bibinfo  {journal} {Phys. Rev. Lett.}\ }\textbf {\bibinfo {volume} {96}},\
  \bibinfo {pages} {111101} (\bibinfo {year} {2006})}\BibitemShut {NoStop}%
\bibitem [{\citenamefont {{Lousto}}\ \emph {et~al.}(2010)\citenamefont
  {{Lousto}}, \citenamefont {{Campanelli}}, \citenamefont {{Zlochower}},\ and\
  \citenamefont {{Nakano}}}]{lousto2010}%
  \BibitemOpen
  \bibfield  {author} {\bibinfo {author} {\bibfnamefont {C.~O.}\ \bibnamefont
  {{Lousto}}}, \bibinfo {author} {\bibfnamefont {M.}~\bibnamefont
  {{Campanelli}}}, \bibinfo {author} {\bibfnamefont {Y.}~\bibnamefont
  {{Zlochower}}}, \ and\ \bibinfo {author} {\bibfnamefont {H.}~\bibnamefont
  {{Nakano}}},\ }\href {\doibase 10.1088/0264-9381/27/11/114006} {\bibfield
  {journal} {\bibinfo  {journal} {Classical Quantum Gravity}\ }\textbf
  {\bibinfo {volume} {27}},\ \bibinfo {eid} {114006} (\bibinfo {year}
  {2010})},\ \Eprint {http://arxiv.org/abs/0904.3541} {arXiv:0904.3541 [gr-qc]}
  \BibitemShut {NoStop}%
\bibitem [{\citenamefont {{Perets}}\ \emph {et~al.}(2016)\citenamefont
  {{Perets}}, \citenamefont {{Li}}, \citenamefont {{Lombardi}},\ and\
  \citenamefont {{Milcarek}}}]{perets2016}%
  \BibitemOpen
  \bibfield  {author} {\bibinfo {author} {\bibfnamefont {H.~B.}\ \bibnamefont
  {{Perets}}}, \bibinfo {author} {\bibfnamefont {Z.}~\bibnamefont {{Li}}},
  \bibinfo {author} {\bibfnamefont {J.~C.}\ \bibnamefont {{Lombardi}},
  \bibfnamefont {Jr.}}, \ and\ \bibinfo {author} {\bibfnamefont {S.~R.}\
  \bibnamefont {{Milcarek}}, \bibfnamefont {Jr.}},\ }\href {\doibase
  10.3847/0004-637X/823/2/113} {\bibfield  {journal} {\bibinfo  {journal}
  {\apj}\ }\textbf {\bibinfo {volume} {823}},\ \bibinfo {eid} {113} (\bibinfo
  {year} {2016})},\ \Eprint {http://arxiv.org/abs/1602.07698} {arXiv:1602.07698
  [astro-ph.HE]} \BibitemShut {NoStop}%
\bibitem [{\citenamefont {{Michaely}}\ \emph {et~al.}(2016)\citenamefont
  {{Michaely}}, \citenamefont {{Ginzburg}},\ and\ \citenamefont
  {{Perets}}}]{michaely2016}%
  \BibitemOpen
  \bibfield  {author} {\bibinfo {author} {\bibfnamefont {E.}~\bibnamefont
  {{Michaely}}}, \bibinfo {author} {\bibfnamefont {D.}~\bibnamefont
  {{Ginzburg}}}, \ and\ \bibinfo {author} {\bibfnamefont {H.~B.}\ \bibnamefont
  {{Perets}}},\ }\href@noop {} {\bibfield  {journal} {\bibinfo  {journal}
  {ArXiv e-prints}\ } (\bibinfo {year} {2016})},\ \Eprint
  {http://arxiv.org/abs/1610.00593} {arXiv:1610.00593 [astro-ph.HE]}
  \BibitemShut {NoStop}%
\bibitem [{\citenamefont {Rosswog}\ \emph {et~al.}(2000)\citenamefont
  {Rosswog}, \citenamefont {Davies}, \citenamefont {Thielemann},\ and\
  \citenamefont {Piran}}]{rosswog_dtp2000}%
  \BibitemOpen
  \bibfield  {author} {\bibinfo {author} {\bibfnamefont {S.}~\bibnamefont
  {Rosswog}}, \bibinfo {author} {\bibfnamefont {M.~B.}\ \bibnamefont {Davies}},
  \bibinfo {author} {\bibfnamefont {F.}~\bibnamefont {Thielemann}}, \ and\
  \bibinfo {author} {\bibfnamefont {T.}~\bibnamefont {Piran}},\ }\href@noop {}
  {\bibfield  {journal} {\bibinfo  {journal} {Astron. Astrophys.}\ }\textbf
  {\bibinfo {volume} {360}},\ \bibinfo {pages} {171} (\bibinfo {year}
  {2000})}\BibitemShut {NoStop}%
\bibitem [{\citenamefont {Lyutikov}(2016)}]{lyutikov2016}%
  \BibitemOpen
  \bibfield  {author} {\bibinfo {author} {\bibfnamefont {M.}~\bibnamefont
  {Lyutikov}},\ }\href@noop {} {\bibfield  {journal} {\bibinfo  {journal}
  {arXiv:1602.07352}\ } (\bibinfo {year} {2016})}\BibitemShut {NoStop}%
\bibitem [{\citenamefont {{Borkovits}}\ \emph {et~al.}(2016)\citenamefont
  {{Borkovits}}, \citenamefont {{Hajdu}}, \citenamefont {{Sztakovics}},
  \citenamefont {{Rappaport}}, \citenamefont {{Levine}}, \citenamefont
  {{B{\'{\i}}r{\'o}}},\ and\ \citenamefont {{Klagyivik}}}]{borkovits2016}%
  \BibitemOpen
  \bibfield  {author} {\bibinfo {author} {\bibfnamefont {T.}~\bibnamefont
  {{Borkovits}}}, \bibinfo {author} {\bibfnamefont {T.}~\bibnamefont
  {{Hajdu}}}, \bibinfo {author} {\bibfnamefont {J.}~\bibnamefont
  {{Sztakovics}}}, \bibinfo {author} {\bibfnamefont {S.}~\bibnamefont
  {{Rappaport}}}, \bibinfo {author} {\bibfnamefont {A.}~\bibnamefont
  {{Levine}}}, \bibinfo {author} {\bibfnamefont {I.~B.}\ \bibnamefont
  {{B{\'{\i}}r{\'o}}}}, \ and\ \bibinfo {author} {\bibfnamefont
  {P.}~\bibnamefont {{Klagyivik}}},\ }\href {\doibase 10.1093/mnras/stv2530}
  {\bibfield  {journal} {\bibinfo  {journal} {Mon. Not. R. Astron. Soc.}\
  }\textbf {\bibinfo {volume} {455}},\ \bibinfo {pages} {4136} (\bibinfo {year}
  {2016})},\ \Eprint {http://arxiv.org/abs/1510.08272} {arXiv:1510.08272
  [astro-ph.SR]} \BibitemShut {NoStop}%
\bibitem [{\citenamefont {{Ransom}}\ and\ \citenamefont
  {et~al.}(2014)}]{ransom2014}%
  \BibitemOpen
  \bibfield  {author} {\bibinfo {author} {\bibfnamefont {S.~M.}\ \bibnamefont
  {{Ransom}}}\ and\ \bibinfo {author} {\bibnamefont {et~al.}},\ }\href
  {\doibase 10.1038/nature12917} {\bibfield  {journal} {\bibinfo  {journal}
  {\nat}\ }\textbf {\bibinfo {volume} {505}},\ \bibinfo {pages} {520} (\bibinfo
  {year} {2014})},\ \Eprint {http://arxiv.org/abs/1401.0535} {arXiv:1401.0535
  [astro-ph.SR]} \BibitemShut {NoStop}%
\bibitem [{\citenamefont {Peters}\ and\ \citenamefont
  {Mathews}(1963)}]{peters_mathews1963}%
  \BibitemOpen
  \bibfield  {author} {\bibinfo {author} {\bibfnamefont {P.~C.}\ \bibnamefont
  {Peters}}\ and\ \bibinfo {author} {\bibfnamefont {J.}~\bibnamefont
  {Mathews}},\ }\href {\doibase 10.1103/PhysRev.131.435} {\bibfield  {journal}
  {\bibinfo  {journal} {Phys. Rev.}\ }\textbf {\bibinfo {volume} {131}},\
  \bibinfo {pages} {435} (\bibinfo {year} {1963})}\BibitemShut {NoStop}%
\bibitem [{\citenamefont {{Eggleton}}\ and\ \citenamefont
  {{Kiseleva}}(1995)}]{eggleton1995}%
  \BibitemOpen
  \bibfield  {author} {\bibinfo {author} {\bibfnamefont {P.}~\bibnamefont
  {{Eggleton}}}\ and\ \bibinfo {author} {\bibfnamefont {L.}~\bibnamefont
  {{Kiseleva}}},\ }\href {\doibase 10.1086/176611} {\bibfield  {journal}
  {\bibinfo  {journal} {\apj}\ }\textbf {\bibinfo {volume} {455}},\ \bibinfo
  {pages} {640} (\bibinfo {year} {1995})}\BibitemShut {NoStop}%
\bibitem [{\citenamefont {{Lousto}}\ and\ \citenamefont
  {{Zlochower}}(2011)}]{zlochower2011}%
  \BibitemOpen
  \bibfield  {author} {\bibinfo {author} {\bibfnamefont {C.~O.}\ \bibnamefont
  {{Lousto}}}\ and\ \bibinfo {author} {\bibfnamefont {Y.}~\bibnamefont
  {{Zlochower}}},\ }\href {\doibase 10.1103/PhysRevLett.107.231102} {\bibfield
  {journal} {\bibinfo  {journal} {Phys. Rev. Lett.}\ }\textbf {\bibinfo
  {volume} {107}},\ \bibinfo {eid} {231102} (\bibinfo {year} {2011})},\ \Eprint
  {http://arxiv.org/abs/1108.2009} {arXiv:1108.2009 [gr-qc]} \BibitemShut
  {NoStop}%
\bibitem [{\citenamefont {{Naoz}}(2016)}]{naoz2016}%
  \BibitemOpen
  \bibfield  {author} {\bibinfo {author} {\bibfnamefont {S.}~\bibnamefont
  {{Naoz}}},\ }\href {\doibase 10.1146/annurev-astro-081915-023315} {\bibfield
  {journal} {\bibinfo  {journal} {Ann. Rev. Astron. Astrophys.}\ }\textbf
  {\bibinfo {volume} {54}},\ \bibinfo {pages} {441} (\bibinfo {year} {2016})},\
  \Eprint {http://arxiv.org/abs/1601.07175} {arXiv:1601.07175 [astro-ph.EP]}
  \BibitemShut {NoStop}%
\bibitem [{\citenamefont {Nissanke}\ \emph {et~al.}(2013)\citenamefont
  {Nissanke}, \citenamefont {Kasliwal},\ and\ \citenamefont
  {Georgieva}}]{nissanke_kg2013}%
  \BibitemOpen
  \bibfield  {author} {\bibinfo {author} {\bibfnamefont {S.}~\bibnamefont
  {Nissanke}}, \bibinfo {author} {\bibfnamefont {M.}~\bibnamefont {Kasliwal}},
  \ and\ \bibinfo {author} {\bibfnamefont {A.}~\bibnamefont {Georgieva}},\
  }\href {\doibase 10.1088/0004-637X/767/2/124} {\bibfield  {journal} {\bibinfo
   {journal} {Astrophys. J.}\ }\textbf {\bibinfo {volume} {767}},\ \bibinfo
  {pages} {124} (\bibinfo {year} {2013})}\BibitemShut {NoStop}%
\bibitem [{\citenamefont {{Meiron}}\ \emph {et~al.}(2016)\citenamefont
  {{Meiron}}, \citenamefont {{Kocsis}},\ and\ \citenamefont
  {{Loeb}}}]{meiron2016}%
  \BibitemOpen
  \bibfield  {author} {\bibinfo {author} {\bibfnamefont {Y.}~\bibnamefont
  {{Meiron}}}, \bibinfo {author} {\bibfnamefont {B.}~\bibnamefont {{Kocsis}}},
  \ and\ \bibinfo {author} {\bibfnamefont {A.}~\bibnamefont {{Loeb}}},\
  }\href@noop {} {\bibfield  {journal} {\bibinfo  {journal} {ArXiv e-prints}\ }
  (\bibinfo {year} {2016})},\ \Eprint {http://arxiv.org/abs/1604.02148}
  {arXiv:1604.02148 [astro-ph.HE]} \BibitemShut {NoStop}%
\bibitem [{\citenamefont {{Seto}}(2008)}]{seto2008}%
  \BibitemOpen
  \bibfield  {author} {\bibinfo {author} {\bibfnamefont {N.}~\bibnamefont
  {{Seto}}},\ }\href {\doibase 10.1086/587785} {\bibfield  {journal} {\bibinfo
  {journal} {Astrophys. J. Lett.}\ }\textbf {\bibinfo {volume} {677}},\
  \bibinfo {eid} {L55} (\bibinfo {year} {2008})},\ \Eprint
  {http://arxiv.org/abs/0802.3411} {arXiv:0802.3411} \BibitemShut {NoStop}%
\bibitem [{\citenamefont {Takahashi}\ and\ \citenamefont
  {Seto}(2002)}]{takahashi_seto2002}%
  \BibitemOpen
  \bibfield  {author} {\bibinfo {author} {\bibfnamefont {R.}~\bibnamefont
  {Takahashi}}\ and\ \bibinfo {author} {\bibfnamefont {N.}~\bibnamefont
  {Seto}},\ }\href {\doibase 10.1086/341483} {\bibfield  {journal} {\bibinfo
  {journal} {Astrophys. J.}\ }\textbf {\bibinfo {volume} {575}},\ \bibinfo
  {pages} {1030} (\bibinfo {year} {2002})}\BibitemShut {NoStop}%
\bibitem [{\citenamefont {Klein}\ \emph {et~al.}(2016)\citenamefont {Klein},
  \citenamefont {Barausse}, \citenamefont {Sesana}, \citenamefont {Petiteau},
  \citenamefont {Berti}, \citenamefont {Babak}, \citenamefont {Gair},
  \citenamefont {Aoudia}, \citenamefont {Hinder}, \citenamefont {Ohme},\ and\
  \citenamefont {Wardell}}]{klein_etal2016}%
  \BibitemOpen
  \bibfield  {author} {\bibinfo {author} {\bibfnamefont {A.}~\bibnamefont
  {Klein}}, \bibinfo {author} {\bibfnamefont {E.}~\bibnamefont {Barausse}},
  \bibinfo {author} {\bibfnamefont {A.}~\bibnamefont {Sesana}}, \bibinfo
  {author} {\bibfnamefont {A.}~\bibnamefont {Petiteau}}, \bibinfo {author}
  {\bibfnamefont {E.}~\bibnamefont {Berti}}, \bibinfo {author} {\bibfnamefont
  {S.}~\bibnamefont {Babak}}, \bibinfo {author} {\bibfnamefont
  {J.}~\bibnamefont {Gair}}, \bibinfo {author} {\bibfnamefont {S.}~\bibnamefont
  {Aoudia}}, \bibinfo {author} {\bibfnamefont {I.}~\bibnamefont {Hinder}},
  \bibinfo {author} {\bibfnamefont {F.}~\bibnamefont {Ohme}}, \ and\ \bibinfo
  {author} {\bibfnamefont {B.}~\bibnamefont {Wardell}},\ }\href {\doibase
  10.1103/PhysRevD.93.024003} {\bibfield  {journal} {\bibinfo  {journal} {Phys.
  Rev. D}\ }\textbf {\bibinfo {volume} {93}},\ \bibinfo {pages} {024003}
  (\bibinfo {year} {2016})}\BibitemShut {NoStop}%
\bibitem [{\citenamefont {Cutler}\ and\ \citenamefont
  {Holz}(2009)}]{cutler_holz2009}%
  \BibitemOpen
  \bibfield  {author} {\bibinfo {author} {\bibfnamefont {C.}~\bibnamefont
  {Cutler}}\ and\ \bibinfo {author} {\bibfnamefont {D.~E.}\ \bibnamefont
  {Holz}},\ }\href {\doibase 10.1103/PhysRevD.80.104009} {\bibfield  {journal}
  {\bibinfo  {journal} {Phys. Rev. D}\ }\textbf {\bibinfo {volume} {80}},\
  \bibinfo {pages} {104009} (\bibinfo {year} {2009})}\BibitemShut {NoStop}%
\bibitem [{\citenamefont {Seto}\ \emph {et~al.}(2001)\citenamefont {Seto},
  \citenamefont {Kawamura},\ and\ \citenamefont {Nakamura}}]{seto_kn2001}%
  \BibitemOpen
  \bibfield  {author} {\bibinfo {author} {\bibfnamefont {N.}~\bibnamefont
  {Seto}}, \bibinfo {author} {\bibfnamefont {S.}~\bibnamefont {Kawamura}}, \
  and\ \bibinfo {author} {\bibfnamefont {T.}~\bibnamefont {Nakamura}},\ }\href
  {\doibase 10.1103/PhysRevLett.87.221103} {\bibfield  {journal} {\bibinfo
  {journal} {Phys. Rev. Lett.}\ }\textbf {\bibinfo {volume} {87}},\ \bibinfo
  {pages} {221103} (\bibinfo {year} {2001})}\BibitemShut {NoStop}%
\bibitem [{\citenamefont {{Poisson}}\ and\ \citenamefont
  {{Will}}(1995)}]{poisson1995}%
  \BibitemOpen
  \bibfield  {author} {\bibinfo {author} {\bibfnamefont {E.}~\bibnamefont
  {{Poisson}}}\ and\ \bibinfo {author} {\bibfnamefont {C.~M.}\ \bibnamefont
  {{Will}}},\ }\href {\doibase 10.1103/PhysRevD.52.848} {\bibfield  {journal}
  {\bibinfo  {journal} {\prd}\ }\textbf {\bibinfo {volume} {52}},\ \bibinfo
  {pages} {848} (\bibinfo {year} {1995})},\ \Eprint
  {http://arxiv.org/abs/gr-qc/9502040} {gr-qc/9502040} \BibitemShut {NoStop}%
\bibitem [{\citenamefont {{Ajith}}(2011)}]{ajith2011}%
  \BibitemOpen
  \bibfield  {author} {\bibinfo {author} {\bibfnamefont {P.}~\bibnamefont
  {{Ajith}}},\ }\href {\doibase 10.1103/PhysRevD.84.084037} {\bibfield
  {journal} {\bibinfo  {journal} {\prd}\ }\textbf {\bibinfo {volume} {84}},\
  \bibinfo {eid} {084037} (\bibinfo {year} {2011})},\ \Eprint
  {http://arxiv.org/abs/1107.1267} {arXiv:1107.1267 [gr-qc]} \BibitemShut
  {NoStop}%
\bibitem [{\citenamefont {{Auchettl}}\ \emph {et~al.}(2016)\citenamefont
  {{Auchettl}}, \citenamefont {{Guillochon}},\ and\ \citenamefont
  {{Ramirez-Ruiz}}}]{auchett2016}%
  \BibitemOpen
  \bibfield  {author} {\bibinfo {author} {\bibfnamefont {K.}~\bibnamefont
  {{Auchettl}}}, \bibinfo {author} {\bibfnamefont {J.}~\bibnamefont
  {{Guillochon}}}, \ and\ \bibinfo {author} {\bibfnamefont {E.}~\bibnamefont
  {{Ramirez-Ruiz}}},\ }\href@noop {} {\bibfield  {journal} {\bibinfo  {journal}
  {ArXiv e-prints}\ } (\bibinfo {year} {2016})},\ \Eprint
  {http://arxiv.org/abs/1611.02291} {arXiv:1611.02291 [astro-ph.HE]}
  \BibitemShut {NoStop}%
\bibitem [{\citenamefont {{Krolik}}\ and\ \citenamefont
  {{Piran}}(2011)}]{krolik2011}%
  \BibitemOpen
  \bibfield  {author} {\bibinfo {author} {\bibfnamefont {J.~H.}\ \bibnamefont
  {{Krolik}}}\ and\ \bibinfo {author} {\bibfnamefont {T.}~\bibnamefont
  {{Piran}}},\ }\href {\doibase 10.1088/0004-637X/743/2/134} {\bibfield
  {journal} {\bibinfo  {journal} {Astrophys. J.}\ }\textbf {\bibinfo {volume}
  {743}},\ \bibinfo {eid} {134} (\bibinfo {year} {2011})},\ \Eprint
  {http://arxiv.org/abs/1106.0923} {arXiv:1106.0923 [astro-ph.HE]} \BibitemShut
  {NoStop}%
\bibitem [{\citenamefont {{Ioka}}\ \emph {et~al.}(2016)\citenamefont {{Ioka}},
  \citenamefont {{Hotokezaka}},\ and\ \citenamefont {{Piran}}}]{ioka2016}%
  \BibitemOpen
  \bibfield  {author} {\bibinfo {author} {\bibfnamefont {K.}~\bibnamefont
  {{Ioka}}}, \bibinfo {author} {\bibfnamefont {K.}~\bibnamefont
  {{Hotokezaka}}}, \ and\ \bibinfo {author} {\bibfnamefont {T.}~\bibnamefont
  {{Piran}}},\ }\href@noop {} {\bibfield  {journal} {\bibinfo  {journal} {ArXiv
  e-prints}\ } (\bibinfo {year} {2016})},\ \Eprint
  {http://arxiv.org/abs/1608.02938} {arXiv:1608.02938 [astro-ph.HE]}
  \BibitemShut {NoStop}%
\bibitem [{\citenamefont {{Stephens}}\ \emph {et~al.}(2011)\citenamefont
  {{Stephens}}, \citenamefont {{East}},\ and\ \citenamefont
  {{Pretorius}}}]{stephens2011}%
  \BibitemOpen
  \bibfield  {author} {\bibinfo {author} {\bibfnamefont {B.~C.}\ \bibnamefont
  {{Stephens}}}, \bibinfo {author} {\bibfnamefont {W.~E.}\ \bibnamefont
  {{East}}}, \ and\ \bibinfo {author} {\bibfnamefont {F.}~\bibnamefont
  {{Pretorius}}},\ }\href {\doibase 10.1088/2041-8205/737/1/L5} {\bibfield
  {journal} {\bibinfo  {journal} {Astrophys. J. Lett.}\ }\textbf {\bibinfo
  {volume} {737}},\ \bibinfo {eid} {L5} (\bibinfo {year} {2011})},\ \Eprint
  {http://arxiv.org/abs/1105.3175} {arXiv:1105.3175 [astro-ph.HE]} \BibitemShut
  {NoStop}%
\bibitem [{\citenamefont {{Rosswog}}\ \emph {et~al.}(2013)\citenamefont
  {{Rosswog}}, \citenamefont {{Piran}},\ and\ \citenamefont
  {{Nakar}}}]{rosswog2013}%
  \BibitemOpen
  \bibfield  {author} {\bibinfo {author} {\bibfnamefont {S.}~\bibnamefont
  {{Rosswog}}}, \bibinfo {author} {\bibfnamefont {T.}~\bibnamefont {{Piran}}},
  \ and\ \bibinfo {author} {\bibfnamefont {E.}~\bibnamefont {{Nakar}}},\ }\href
  {\doibase 10.1093/mnras/sts708} {\bibfield  {journal} {\bibinfo  {journal}
  {Mon. Not. R. Astron. Soc.}\ }\textbf {\bibinfo {volume} {430}},\ \bibinfo
  {pages} {2585} (\bibinfo {year} {2013})},\ \Eprint
  {http://arxiv.org/abs/1204.6240} {arXiv:1204.6240 [astro-ph.HE]} \BibitemShut
  {NoStop}%
\end{thebibliography}

%
\end{document}